\documentclass[12pt]{article}
\usepackage{epsfig}
\usepackage{amsfonts}
\usepackage{amsopn}
\usepackage{amsmath}

\title{
\vskip -50 pt
\begin{flushright}
\normalsize\rm AEI-2012-051 \\ 
\end{flushright}
\vskip 20 pt
A note on  the principle of least action and Dirac matrices}

\author{
Maciej Trzetrzelewski \thanks{e-mail: maciej.trzetrzelewski@gmail.com} \\ \\
Max-Planck-Institut f\"ur Gravitationsphysik, \\
Albert-Einstein-Institut, \\
M\"uhlenberg 1, D-14476 Potsdam, \\
Germany}

\begin{document}
\date{}
\maketitle

\abstract{Many Lagrangians of physical theories can be expressed as eigenvalues of certain, relatively simple,  matrices involving Dirac gamma matrices. We give concrete examples for Lagrangian corresponding to  a point particle coupled to electromagnetic field, electrodynamics, nonabelian gauge theories, extended objects and gravity.  We also discuss (in case of a point particle) what are the implications of  the least action principle applied to matrix Lagrangians.}

\section{Motivation}

A nice property of  Dirac matrices in 4 dimensions is that a large class of Lagrangians appearing in classical physics can be understood as eigenvalues of some matrices  involving them. As for the Lagrangian for a point particle this observation is almost trivial considering the context in which Dirac introduced gamma matrices - it also works in arbitrary spacetime dimension. In the following sections we would like to note some less trivial observations, in particular that the Lagrangians of 4D gauge theories and general relativity as well as extended objects can also be represented as eigenvalues of certain matrices.  Lastly we consider the variation procedure for actions that are matrices (not eigenvalues of matrices) and discuss conditions under which one recovers classical equations of motion. 

The fact that Dirac matrices appear so naturally at the level of classical theories (although they originate from Dirac's attempt to formulate relativistically invariant \emph{quantum} theory) is somewhat intriguing.  If not purely accidental and formal it must be a consequence of some underlying physical reason. For the letter a good explanation would a claim that the spin (and hence the Dirac/spin matrices) is the property of space-time itself which is then inherited by particles. Under these circumstances it would be natural to expect that relativistic field theories involve (in a certain way) Dirac matrices already at the classical level.

\section{Lagrangians as eigenvalues}

\subsection{Point particles}
For a point particle with mass $m$, charge $q$ and coordinates $x^{\mu}$, coupled to the electromagnetic field $A_{\mu}(x)$, in curved spacetime given by the metric $g^{\mu\nu}(x)$, one simply takes 
\begin{equation} \label{point}
P:=m\gamma_{\mu}(x)\dot{x}^{\mu}+\bold{1}q A_{\mu}(x)\dot{x}^{\mu}, \ \ \ \{\gamma_{\mu}(x),\gamma_{\nu}(x)\}=2g_{\mu\nu}(x).  
\end{equation}
Here $\gamma^{\mu}(x)$ are $x$ dependent Dirac matrices. One possible choice is to use the vielbein $e^a_{\mu}(x)$, $g_{\mu\nu}(x)= e_{\mu}^a(x) e_{\nu}^b(x) \eta_{ab}$ where $\eta_{ab}$ is the Minkowski metric (we use the many minus convention when needed) and to take   $\gamma_{\mu}(x)= e_{\mu}^a(x) \gamma_a $ where $\gamma_a$ are usual ($x$ independent) Dirac matrices $\{\gamma_a,\gamma_b\}=2\eta_{ab}$. 

Because the square of the mass term of $P$ is just $m^2g_{\mu\nu}(x)\dot{x}^{\mu}\dot{x}^{\nu}$  and since  the potential term commutes with the mass term
we conclude that there are two doubly degenerated eigenvalues of $P$ and the spectrum is
\[
spec(P)=\{P_{+},P_{+},P_{-},P_{-}\}, \ \ \ P_{\pm}=\pm m\sqrt{g_{\mu\nu}(x)\dot{x}^{\mu}\dot{x}^{\nu}} +q A_{\mu}(x)\dot{x}^{\mu}.
\]
Hence the Lagrangian for a point-particle can be obtained by taking the smallest eigenvalue, $P_{-}$, of $P$ so that the least action principle gives the correct equations of motion.

\subsection{Gauge theories}
A similar result may be obtained for a Lagrangian of electrodynamics in flat space in 4 dimensions. One takes
\begin{equation} \label{ed}
E:=\frac{1}{8}(\gamma^{ab}F_{ab})^2 =-\frac{1}{4}F_{ab}F^{ab}\bold{1}-\frac{1}{4} i \gamma^{5}F_{ab} \  ^*F^{ab}
\end{equation}
where $\gamma^{ab}:=[\gamma^a,\gamma^b]/2$, $F_{ab}:=\partial_{[a}A_{b]}$ is the field strength and the dual tensor is $^*F^{ab}:=\frac{1}{2}\epsilon^{abcd}F_{cd}$ where $\epsilon^{abcd}$ is completely antisymmetric with $\epsilon^{0123}=1$. The second equality in (\ref{ed}) is unique for 4-dimensional spacetimes and  relies on the following identity for gamma matrices
\begin{equation} \label{iden}
\{\gamma^{ab},\gamma^{cd}\} =-2 (\eta^{ac}\eta^{bd}-\eta^{ad}\eta^{bc} )\bold{1}-2i \gamma^5 \epsilon^{abcd}.
\end{equation}
The eigenvalues of $E$ are again double degenerated and its spectrum is
\begin{equation} \label{speced}
spec(E)=\{E_{+},E_{+},E_{-},E_{-}\}, \ \ \ E_{\pm}=-\frac{1}{4} F_{ab}F^{ab}\pm  \frac{1}{4} i F_{ab} \  ^*F^{ab}.
\end{equation}
(where we use the fact that there are two doubly degenerated eigenvalues of a matrix $a+\gamma^5 b$: $a\pm b$).
The second term in $E_{\pm}$ is a total derivative and hence does not affect the equations of motions therefore to obtain the formulation of electrodynamics one can take $E_{+}$ or $E_{-}$\footnote{One can also take $^*F_{ab}$ instead of $F_{ab}$ in the definition of $E$ obtaining equivalent result.}. Before going further let us note that by  equation  (\ref{ed})
it is not possible to couple a pseudo-scalar field (the axion) to the $F^*F$ term, leaving the $FF$ term unchanged. Therefore from this perspective the coupling to the axion field does not appear here quite naturally.

For non-abelian gauge theories we may take
\[
E_{YM}:=-\frac{1}{4} (\gamma^{ab}G_{ab}^A)^2= -\frac{1}{4}G^A_{ab}G^{ab \ A}\bold{1}-\frac{1}{4} i \gamma^{5}G^A_{ab} \  ^*G^{ab \ A}
\]
where  $A$ is the adjoint index of the gauge group and $G^A_{ab}$ is the $A$'th component of the field strength (reducing to $F_{ab}$ if the group is $U(1)$). Here the conclusions are similar to (\ref{speced}) with one difference that the $G^*G$ term must be kept due to the presence of instantons. Therefore the $\theta$ angle is fixed and equal to $\theta=8\pi^2$ (using the conventions where the $\theta$ term is given by$ -\frac{\theta i }{32\pi^2}G^A_{ab} \  ^*G^{ab \ A}  $). 

\subsection{General Relativity}
For the Lagrangian of General Relativity we find that a good choice is
\begin{equation} \label{gr}
G:=\frac{1}{4}R_{\mu\nu\rho\sigma}\gamma^{\mu\nu}(x)\gamma^{\rho\sigma}(x)= R\mathbf{1} -2i \gamma^5 \ ^*R = R\mathbf{1}.
\end{equation}
where $\gamma^{\mu\nu}(x)=e^{\mu}_a e_b^{\nu} \gamma^{ab}$ and $^*R$ is a dual to the Riemann tensor. To prove that  (\ref{gr}) holds one again uses (\ref{iden}). Moreover, in $4$ dimensions that dual $^*R$ is zero due to the Bianchi identities
\[
^*R:=\epsilon^{\mu\nu\rho\sigma}R_{\mu\nu\rho\sigma}=8(R_{0123}+R_{0231}+R_{0312})=0
\]
therefore matrix $G$ is actually a unit matrix proportional to the Ricci scalar. 

\subsection{Extended objects}
 For extended objects (with $p$ spatial dimensions) the task is more tricky since the corresponding Lagrangians are more complicated (compared to a point-like particle). Their Lagrangians are given by \cite{Dirac,Nambu,Goto}
\[
L=-\Lambda \sqrt{(-1)^p\det G_{\alpha\beta}}, \ \ \ \ G_{\alpha\beta}= \partial_{\alpha}X^{\mu}\partial_{\beta}X^{\nu}g_{\mu\nu}(X)
\]  
and the action is obtained by integrating $L$ over the world-volume parametrized by $\sigma^{\alpha}$, $\alpha=0,\ldots,p$ (for strings, $p=1$, for membranes, $p=2$). Here $X^{\mu}$ are the coordinates of the object  and $g_{\mu\nu}(X)$ is the metric of spacetime. Dimensional constant $\Lambda$ is the tension. Using observations made in \cite{33lewski} we find that a good choice of the matrix Lagrangian is
\begin{equation} \label{string}
P_{string}=\frac{\Lambda}{2!} \gamma_{\mu \nu}(X)\{X^{\mu},X^{\nu} \},
\end{equation}
\[
\{X^{\mu},X^{\nu} \} = \epsilon^{\alpha\beta}\partial_{\alpha}X^{\mu}\partial_{\beta}X^{\nu}
\]
for a string and
\begin{equation} \label{membrane}
P_{membrane}=\frac{\Lambda}{3!} \gamma_{\mu \nu \rho}(X)\{X^{\mu},X^{\nu}, X^{\rho} \}, 
\end{equation}
\[
\{X^{\mu},X^{\nu}, X^{\rho} \} =  \epsilon^{\alpha\beta \gamma}\partial_{\alpha}X^{\mu}\partial_{\beta}X^{\nu}\partial_{\gamma}X^{\rho}
\]
for a membrane, where $\gamma^{\mu\nu\rho}(X)=e^{\mu}_a e^{\nu}_b e^{\rho}_c\gamma^{abc}$, $\gamma^{abc}=\frac{1}{3}(\gamma^{\mu\nu}\gamma^{\rho}+cycl.)$. We note a natural appearance of the Poisson and Nambu brackets in (\ref{string}) and (\ref{membrane}) respectively. For higher dimensional objects $p>2$ (which would require the embedding spacetime to have more dimensions then 4) one simply takes $\frac{\Lambda}{(p+1)!} \gamma_{\mu_1 \ldots \mu_{p+1}}\{X^{\mu},\ldots, X^{\mu_{p+1}} \}$. 

Let us now argue why the eigenvalues of matrices (\ref{string}) and (\ref{membrane}) give the corresponding Lagrangians. Using identity (\ref{iden}) one immediately verifies that the square of $P_{string}$ is equal to $-\det G_{\alpha\beta}$, $\alpha,\beta=0,1$. To prove the same result for a membrane it is useful to use another identity
\[
\{\gamma^{abc},\gamma^{a'b'c'}\} = 2\epsilon^{abcd}\epsilon^{a'b'c'd'}\eta_{dd'}\mathbf{1}= -
2\left| \begin{array}{ccc}
                     \eta^{aa'}  & \eta^{ab'} & \eta^{ac'}   \\
                      \eta^{ba'}  & \eta^{bb'} & \eta^{bc'}   \\
                        \eta^{ca'}  & \eta^{cb'} & \eta^{cc'}  
                              \end{array} \right| \bold{1}
\] 
which follows form the fact that in 4 dimensions $\gamma^{abc}=i\epsilon^{abcd}\gamma_d \gamma^5$. Now it is easy to see that the square of $P_{membrane}$ is equal to $\det G_{\alpha\beta}$, $\alpha,\beta=0,1,2$.

\section{Action principle for matrices?}
Let us apply the least action principle directly to a matrix Lagrangian (not its eigenvalue) of a neutral particle in curved space
\begin{equation} \label{action}
S=m\int \gamma_{\mu}(x)\dot{x}^{\mu}d\tau.
\end{equation}
In the case of a flat space the matrix Lagrangian in $S$ is a total derivative hence  $\delta S=0$ for all trajectories - which is of course inconsistent with the geodesic equation.

In curved space the conclusion is different. Varying $S$ with respect to $x^{\mu}$ and integrating by parts we obtain  
\begin{equation}  \label{var}
\delta S = \int (\dot{x}^{\mu}\partial_{\nu}\gamma_{\mu}(x)-\dot{\gamma}_{\nu})\delta x^{\nu}d\tau
\end{equation}
hence the equations of motion are $\dot{x}^{\mu}\partial_{\nu}\gamma_{\mu}(x)=\dot{\gamma}_{\nu}$ which can be also written as
\begin{equation} \label{eomp}
\dot{x}^{\mu}f_{\mu\nu}=0, \ \ \ \ f_{\mu\nu}:=\partial_{[\mu}\gamma_{\nu]}(x).
\end{equation}
However the only solution of this equation is $\dot{x}^{\mu}=0$ which can be seen by  expanding the matrix $f_{\mu\nu}$ in terms of $\gamma_a$'s  as
\[
f_{\mu\nu}= f_{\mu\nu}^a\gamma_a, \ \ \ \ f_{\mu\nu}^a:= \partial_{[\mu}e^a_{\nu]}
\]
therefore equation (\ref{eomp}) implies $\dot{x}^{\mu}f^a_{\mu\nu}=0$ for every $a$ hence $\dot{x}^{\mu}=0$. This shows that the direct application of the least action principle to matrices via (\ref{action}) is inconsistent with the geodesic equation.

In the remaining part of this section we argue that the geodesic equation can nevertheless be obtained for certain geometries if the equations of motion (\ref{eomp}) are  modified as follows
\begin{equation}  \label{eommod}
\dot{x}^{\mu}f_{\mu\nu}=C_{\nu}.
\end{equation}
Here $C_{\nu}$ is some constant, nonzero matrix and can be obtained by evaluating the l.h.s. of (\ref{eommod}) at some time $\tau_0$. Therefore matrix $C_{\nu}$ is completely determined by initial conditions of the trajectory $x^{\mu}(\tau)$ via $C_{\nu}=\dot{x}^{\mu}f_{\mu\nu}|_{\tau=\tau_0}$. This modification could be argued from equation (\ref{var}) if we generalize the variation procedure by saying that the integrand in (\ref{var}) is at most a  traceless (but nor necessarily 0) matrix. The matrix $C_{\nu}$ can be expanded in terms of $\gamma_a$'s and therefore satisfies this condition.

Let us now find solutions of (\ref{eommod}). Ideally, we would like to find the inverse of $f_{\mu\nu}$ so that equation (\ref{eommod}) could be inverted. This can be done by noting that in 4 dimensions there exist a unique identity for antisymmetric rank-2 tensors and their duals, namely \cite{Bjorken}
\[
f^a_{\mu\nu}\ ^*f^{\nu\rho \ a} =-(f^a_{\nu\sigma}\ ^*f^{\nu\sigma \ a})\delta^{\rho}_{\mu}, \ \ \ \ \hbox{$a$ fixed},
\]
where $^*f^{\mu\nu \ a}$ is a dual tensor $ ^*f^{\mu\nu \ a}:=\epsilon^{\mu\nu\rho\sigma}f^a_{\rho\sigma}/2$. This identity  implies that 
\begin{equation} \label{fdual}
\{f_{\mu\nu},\ ^*f^{\nu\rho}\}=-2\delta^{\rho}_{\mu}(f^*f), \ \ \ \ (f^*f):=f^a_{\nu\sigma}\ ^*f^{\nu\sigma \ b}\eta_{ab},
\end{equation}
where $^*f^{\mu\nu}$ is a dual matrix $^*f^{\mu\nu}:=^*f^{\mu\nu \ a}\gamma_a$. Applying (\ref{fdual}) to (\ref{eommod}) we find that
\begin{equation} \label{solmod}
\dot{x}^{\rho}=-\frac{1}{2(f^*f)} \{C_{\nu},^*f^{\nu\rho}\},
\end{equation}
where we assumed that the geometry under consideration is such that $(f^*f)\ne 0$. That assumption implies that the metric $g_{\mu\nu}(x)$ is non standard (i.e. not diagonal). To see this explicitly let us assume that the metric is given by $g_{\mu\nu}(x)=\eta_{\mu\nu}f^2_{\mu}(x)$ ($\mu$ fixed) hence $e^a_{\mu}=\delta^a_{\mu}f_{\mu}(x)$. In this case it follows that $(f^*f)$ is
\[
(f^*f)=[-(\partial_{0-1}f_{\mu}) (\partial_{2-3}f_{\nu})+ (\partial_{0-2}f_{\mu}) (\partial_{1-3}f_{\nu})  -(\partial_{0-3}f_{\mu}) (\partial_{1-2}f_{\nu})] \eta^{\mu\nu}
\]
(where $\partial_{a-b}:=\partial_a-\partial_b$) which is $0$. Therefore equation (\ref{solmod}) applies to very special geometries. 

Equation (\ref{solmod}) is an autonomic one i.e. it  gives the formula for the 4-velocities $\dot{x}^{\mu}$ in terms of $x^{\mu}$'s. To obtain the solutions for $x^{\mu}$ one would have to integrate (\ref{solmod}) once, which however can be done explicitly only when the metric is specified.

Let us now show that $x^{\mu}$ satisfying  (\ref{eommod}) satisfies also the geodesic equation. To see this it is useful to differentiate (modified) equations of motion (\ref{eommod}) w.r.t. $\tau$. We find that
\[
\ddot{x}^{\mu}f_{\mu\nu}+\dot{x}^{\mu}\dot{x}^{\sigma}\partial_{\sigma}f_{\mu\nu}=0.
\] 
Applying (\ref{fdual}) to the above equation we obtain
\[
\ddot{x}^{\rho}+\tilde{\Gamma}^{\rho}_{\mu\sigma} \dot{x}^{\mu}\dot{x}^{\sigma}=0, \ \ \ \tilde{\Gamma}^{\rho}_{\mu\sigma}:= \frac{1}{2(f^*f)} \ ^*f^{\rho\nu \ a}\partial_{(\sigma}f_{\mu)\nu}^b \eta_{ab}.
\]
Therefore we arrive at the geodesic equation provided $\tilde{\Gamma}^{\rho}_{\mu\sigma}$ coincide with Christoffel symbols $\Gamma^{\rho}_{\mu\sigma}$.

\section{Acknowledgements}
This work was supported by DFG (German Science Foundation) via
the SFB grant.

\end{document}